\documentclass[11pt]{article}

\usepackage[letterpaper,margin=1in]{geometry}
\usepackage[T1]{fontenc}
\usepackage{lmodern}
\usepackage{amsmath,amssymb}
\usepackage{booktabs,longtable,array,calc}
\usepackage{graphicx}
\usepackage[section]{placeins}
\usepackage{microtype}
\usepackage{hyperref}
\usepackage{xurl}
\usepackage{etoolbox}
\hypersetup{
  hidelinks,
  pdftitle={Paging the Experts: A Reproducible Characterization of Flash-Backed MoE Inference on iPhone},
  pdfauthor={Musa Shams},
  pdfsubject={Flash-backed mixture-of-experts inference on iPhone},
  pdfkeywords={on-device inference, mixture-of-experts, expert caching, performance characterization, reproducibility}
}
\setkeys{Gin}{width=\linewidth,keepaspectratio}
\AtBeginEnvironment{longtable}{\small\setlength{\tabcolsep}{4pt}}
\providecommand{\tightlist}{%
  \setlength{\itemsep}{0pt}\setlength{\parskip}{0pt}}

\title{Paging the Experts: A Reproducible Characterization of Flash-Backed MoE Inference on iPhone}
\author{Musa Shams\\
\textit{Independent Researcher}\\
\href{https://orcid.org/0009-0005-1015-5342}{ORCID: 0009-0005-1015-5342}}
\date{}

\begin{document}
\maketitle

\begin{abstract}
Sparse activation reduces mixture-of-experts computation without eliminating
the need to store all experts. We present Routide, a Swift/MLX runtime that
executes the text path of a pinned public Qwen3.6-35B-A3B quantized checkpoint
while keeping expert weights in iPhone storage and a byte-budgeted subset in
memory. We characterize cache-policy sensitivity, numerical comparison
boundaries, and measurement limits. Across five recorded 128-token
workloads, fixed-route replay gives 0.00\% demand hits with a 512 MiB LRU
cache, 18.80\% with seeded random eviction at the same budget, and
38.58\% with 576 MiB LRU. The apparent capacity cliff is therefore a
policy/workload interaction, not a universal memory requirement. Same-runtime Mac controls preserve generated sequences across eviction and asynchronous
prefetch, including 2,560 exact token comparisons and 10,334
speculative loads. In contrast, complete resident-Python versus recorded-phone
sequences disagree on all five tested cases, precluding a general numerical
equivalence claim. Two separately scoped iOS 27 memory protocols observe
sampled process-footprint peaks of 1.87-2.32 GiB
on short prompts and 2.39-2.73 GiB on one
longer prompt. We retain a thermal stopping event, negative timing comparisons,
and a single qualified whole-device power estimate. These results establish
bounded feasibility and identify limitations that a deployment claim must not
hide.
\end{abstract}

\hypertarget{introduction}{%
\section{Introduction}\label{introduction}}

An MoE model can have a modest active computation graph per token while its
full weight set remains too large for a phone's memory. Quantization reduces
this gap but does not close it for every model/device combination. Storage-backed
execution is consequently a systems problem involving layout, cache admission
and eviction, I/O scheduling, framework arithmetic, and measurement semantics.
A demonstration that emits text does not by itself establish efficient
execution, numerical equivalence, or improved task quality.

Routide uses a publicly identified checkpoint:
\texttt{mlx-community/Qwen3.6-35B-A3B-4bit}, revision
\texttt{38740b847e4cb78f352aba30aa41c76e08e6eb46} {[}1{]}. The model's nominal description
is approximately 35 billion total parameters and 3 billion active parameters
per token. We do not treat those rounded architectural labels as measured
memory requirements. Instead, we inventory actual tensors, count expert
payload bytes, and separately sample process memory.

We make three contributions:

\begin{enumerate}
\def\labelenumi{\arabic{enumi}.}
\tightlist
\item
  \textbf{An inspectable storage-to-execution path.} A bounded-memory packer,
  aligned per-expert layout, Swift positional reader, byte-budgeted cache,
  quantized MLX executor, and on-device benchmark capture form a complete
  implementation for this checkpoint's text path.
\item
  \textbf{A cache-policy characterization with informative negative results.}
  Replays explain why naive LRU thrashes at one budget, demonstrate that
  the behavior is not inevitable, and expose resident-prefetch checks as
  a source of cache pollution even when no speculative read occurs.
\item
  \textbf{Separated correctness and measurement claims.} We distinguish
  same-runtime cache transparency from cross-runtime sequence equivalence,
  logical expert traffic from physical storage traffic, and sampled process
  footprint from MLX allocator counters. A stopped phone campaign and its
  separately declared continuation remain separate evidence.
\end{enumerate}

We do not claim the first phone LLM runtime, state-of-the-art throughput,
quality parity with a resident model, or prefetch energy savings. The
experiments support a narrower but reproducible characterization.

\hypertarget{background-and-related-work}{%
\section{Background and related work}\label{background-and-related-work}}

\textbf{Flash-backed inference.} \emph{LLM in a flash} studies models larger than available
DRAM and emphasizes reducing transferred data and using larger contiguous
reads. Its windowing and row-column bundling exploit activation sparsity {[}2{]}.
Routide follows the storage-aware motivation but pages explicitly routed,
quantized MoE experts rather than introducing a learned neuron-sparsity
predictor or modifying activation functions. We do not reproduce that paper's
physical-storage bandwidth experiments or compare its reported speed directly
with our phone timings.

\textbf{Phone-oriented sparse execution.} PowerInfer-2 combines neuron-granularity
caching, heterogeneous execution, and fine-grained I/O/computation pipelining
on smartphones {[}3{]}. Its model and hardware choices differ from ours.
Routide's scope is a fixed public checkpoint on Apple hardware, including
byte-accounting and numerical controls; it does not implement PowerInfer-2's
planner or establish superiority over it.

\textbf{MoE offloading.} Fiddler reduces CPU-GPU data movement by orchestrating expert
execution across those processors {[}4{]}. Its discrete CPU/GPU memory hierarchy
differs from iPhone unified memory plus storage. Both settings illustrate why
total parameter size, active arithmetic, and data movement must be considered
separately.

\textbf{Runtime and cache references.} Routide builds on Apple's MLX Swift ecosystem
and the \texttt{LLMEval} application example {[}5{]}. Farthest-next-use replacement is a
classical offline paging reference {[}6{]}. Here we distinguish a feasible
pin-constrained schedule from a relaxed, unpinned equal-block oracle.
Neither can serve as an online predictor without access to future routing.

\hypertarget{system-design}{%
\section{System design}\label{system-design}}

\hypertarget{checkpoint-and-storage-inventory}{%
\subsection{Checkpoint and storage inventory}\label{checkpoint-and-storage-inventory}}

We use binary MiB/GiB (\texttt{2\^{}20}/\texttt{2\^{}30} bytes) and decimal GB (\texttt{10\^{}9} bytes).

The pinned checkpoint identifies its architecture as \texttt{qwen3\_5\_moe}. The text
path has 40 layers, 256 routed experts per layer, and top-eight expert selection.
Ten layers use full attention and 30 use recurrent gated-delta attention.
The checkpoint is primarily affine INT4 with group size 64, with higher-precision
router-related overrides and BF16 quantization metadata; ``4-bit'' does not
mean every stored scalar is four bits.

\begin{longtable}[]{@{}
  >{\raggedright\arraybackslash}p{(\columnwidth - 4\tabcolsep) * \real{0.3000}}
  >{\raggedleft\arraybackslash}p{(\columnwidth - 4\tabcolsep) * \real{0.4000}}
  >{\raggedright\arraybackslash}p{(\columnwidth - 4\tabcolsep) * \real{0.3000}}@{}}
\toprule\noalign{}
\begin{minipage}[b]{\linewidth}\raggedright
Quantity
\end{minipage} & \begin{minipage}[b]{\linewidth}\raggedleft
Bytes
\end{minipage} & \begin{minipage}[b]{\linewidth}\raggedright
Interpretation
\end{minipage} \\
\midrule\noalign{}
\endhead
\bottomrule\noalign{}
\endlastfoot
Checkpoint tensor payload & 20,401,929,952 & Includes the vision tower \\
Text-only resident-reference parameters & 19,508,787,456 & Retained language tensors, not process RAM \\
Excluded vision tensors & 893,142,496 & Explicitly excluded by the text loader \\
Routed-expert payload & 18,119,393,280 & All layers and experts \\
Non-expert checkpoint payload & 2,282,536,672 & Not an always-resident runtime set \\
Complete packed files & 20,403,825,982 & Includes manifest and alignment \\
One routed-expert payload & 1,769,472 & 1.6875 MiB; 27 aligned 64 KiB units \\
\end{longtable}

The fully resident Python text loader retains 1,757 tensors and deliberately
excludes 333 vision tensors. We verify retained names, shapes, types, and byte
counts, not merely a smaller-than-expected total. The inventory difference is
not evidence that language-model weights were discarded.

The pack comprises \texttt{manifest.json}, \texttt{resident.bin}, and one expert file per
layer. Each expert block stores its gate, up, and down projections together
with scales and biases; blocks start on 64 KiB boundaries. The format name
\texttt{resident.bin} means the \textbf{non-routed partition of the checkpoint}, not a
promise to load every byte into RAM. The text runtime loads the required
attention, normalization, router, shared-expert, and output-head tensors,
does not execute the vision tower, and streams the requested quantized input
embedding row. Thus, packed-file size, text parameter size, and measured
process footprint are different quantities.

\hypertarget{demand-execution-and-cache-semantics}{%
\subsection{Demand execution and cache semantics}\label{demand-execution-and-cache-semantics}}

The range reader uses positional reads rather than mapping the full model.
The cache is keyed by \texttt{(layer,\ expert)} and budgets expert payload bytes.
Entries are pinned while a layer evaluates them; entries selected earlier in
the same route remain protected until that layer's operation completes.
Weights remain packed quantized arrays during projection, rather than being
expanded into a full FP16 expert copy. Activations, quantization metadata,
framework allocations, staging data, attention state, and OS behavior still
contribute to memory beyond the expert payload budget.

For a single forward, 40 layers each request eight distinct routed experts:
320 requests and 540 MiB of expert payload \textbf{if all miss}. This is an
over-the-forward working set, not 540 MiB of simultaneously pinned experts.
The byte budget and pin lifetime are modeled in the simulator. Bypass
accounting exists for infeasible admissions; the reported replay grid has
zero oversized-expert or pinned-working-set bypasses.

Routide performs serial, single-token prefill and greedy generation. If a
request has \texttt{P} prepared prompt tokens and \texttt{G} sampled output tokens, with
at least one output, it performs \texttt{P\ +\ G\ -\ 1} forwards. It computes final
logits for each serial input token; batched prefill is not implemented.
Reported TTFT therefore characterizes this implementation, not an optimized
batched-prefill baseline.

\hypertarget{prefetch-as-a-separate-optional-policy}{%
\subsection{Prefetch as a separate, optional policy}\label{prefetch-as-a-separate-optional-policy}}

The tested confidence-gated predictor selects the preceding step's
highest-weight expert \textbf{in the same layer}, if its normalized selected-expert
weight is at least 0.20. It schedules a nonblocking request before that layer's
attention computation during prefill only. This is not a learned predictor
of the next layer. Actual router selection and expert arithmetic are unchanged.
The default remains \textbf{576 MiB / LRU / None}.

The original resident-hit behavior refreshed recency when a prediction was
already cached. The separate preserve-resident policy records the probe
without changing recency, frequency, pins, or access clock. Missing experts
use the same asynchronous load and accounting path. A policy can therefore
have many prediction checks but no new I/O; this distinction is essential
when interpreting apparent prefetch benefits.

\hypertarget{experimental-methodology}{%
\section{Experimental methodology}\label{experimental-methodology}}

\hypertarget{evidence-cohorts-and-scope}{%
\subsection{Evidence cohorts and scope}\label{evidence-cohorts-and-scope}}

The unit of evidence is the recorded protocol/build, not an interchangeable
``Routide run.'' We retain different operating systems, runtime versions,
instrumentation, prompts, and generation caps rather than pooling them.

\begin{longtable}[]{@{}
  >{\raggedright\arraybackslash}p{(\columnwidth - 6\tabcolsep) * \real{0.2500}}
  >{\raggedright\arraybackslash}p{(\columnwidth - 6\tabcolsep) * \real{0.2500}}
  >{\raggedright\arraybackslash}p{(\columnwidth - 6\tabcolsep) * \real{0.2500}}
  >{\raggedright\arraybackslash}p{(\columnwidth - 6\tabcolsep) * \real{0.2500}}@{}}
\toprule\noalign{}
\begin{minipage}[b]{\linewidth}\raggedright
Cohort
\end{minipage} & \begin{minipage}[b]{\linewidth}\raggedright
Workload and purpose
\end{minipage} & \begin{minipage}[b]{\linewidth}\raggedright
Platform
\end{minipage} & \begin{minipage}[b]{\linewidth}\raggedright
Scope
\end{minipage} \\
\midrule\noalign{}
\endhead
\bottomrule\noalign{}
\endlastfoot
Initial capacity study & One-character prompt (\texttt{.}); 23 prepared tokens, 8 outputs; five cold/warm repetitions per budget & iPhone 17 Pro Max, iOS 26.5.2 & Narrow LRU calibration \\
Held-out routes & Five original \texttt{-001} prompts, 128 outputs each & Same phone, iOS 26.6.1 & Measured routes; offline policy/prefetch replay \\
Counting diagnostics & Six balanced 128-token pairs; separately, four 512-token ABBA runs & Same phone, iOS 26.6.1 & Exploratory timing, traffic, sustained feasibility \\
Resident reference & Exact phone input IDs; five \texttt{-001} cases & M1 Ultra, 128 GiB, macOS 26.6.2; Python MLX 0.31.2 & Cross-runtime diagnostic, not speed \\
Same-runtime controls & Five \texttt{-001} cases; five additional \texttt{-003} cases & Actual Swift model on the same high-memory Mac & Eviction/prefetch transparency \\
Process-memory parent & Three \texttt{-001} prompts, two budget pairs each; stopped on request 12 & Phone, iOS 27.0; executable \texttt{9da46f6e...} & Twelve retained rows, eleven fully valid \\
Longer-context follow-up & One fixed longer prompt, one request per budget & Phone, iOS 27.0; executable \texttt{fb4a9a64...} & Separate two-request memory-only protocol \\
Power case study & One profiler-recorded 512-token counting request & Phone, iOS 26.6.1 & Whole-device source-window estimate \\
\end{longtable}

The five \texttt{-001} prompts cover conversation, code, mathematics, reasoning, and
expository text. They differ from the \texttt{-002} prompts used in an earlier live
timing suite and the counting workload used to choose the confidence
threshold. The \texttt{-003} prompts provide an additional fixed set for the final
Mac correctness matrix. These are small designed workloads, not independent
samples from a representative production distribution or scored quality tests.

The cache-policy sensitivity grid was frozen \textbf{after the route captures were
available but before replaying the new grid}. It contains 448, 512, 576, 640,
768, and 1,024 MiB budgets, six pin-constrained policies, and an additional
relaxed oracle. Random uses all five declared seeds, 0-4. The 300 pinned
replays and 30 relaxed bounds are correlated counterfactuals over five traces,
not 330 independent model runs.

\hypertarget{counters-timestamps-and-statistics}{%
\subsection{Counters, timestamps, and statistics}\label{counters-timestamps-and-statistics}}

\texttt{expertBytesRead} counts logical packed \textbf{expert} payload delivered by the
reader. It does not count resident-model preparation or per-token embedding
row reads, and is not a measurement of physical NAND traffic, storage-device
wear, or all system I/O. With no speculation it equals demand misses times
the expert payload size. With speculation, new speculative loads contribute
their bytes; demand joins and useful/wasted classifications are retained
separately. A cold expert cache does not imply a cold filesystem cache.

We report measured elapsed, TTFT, and decode throughput in their saved
scopes. Paged decode throughput uses the post-first-token interval; request
timing also records metric/prefetch drain. Model/tokenizer preparation is
included when it occurs inside a request. Automated preloads and warm-ups
must not be mistaken for measured model-load time. Precise device Unix
timestamps are cross-checked against a monotonic duration. Copy/export
timestamps are never substituted for request boundaries.

For policy replay, hit rates are request-weighted across all five prompts;
cache state continues from prefill into decode. Random results use the mean
and range of the five full-suite seeds, \textbf{not confidence intervals}.
For the small live paired study we retain every pair, including an opposite
direction and the pair containing the thermal violation. We report
descriptive quantities, not p-values or population-level confidence claims.

\hypertarget{memory-and-environmental-observation}{%
\subsection{Memory and environmental observation}\label{memory-and-environmental-observation}}

Schema-12 phone benchmarks sample \texttt{TASK\_VM\_INFO.resident\_size} and
\texttt{phys\_footprint} on a separate serial queue every 250 ms, with start/end
attempts and an actual maximum successful-sample gap. They also record
process dirty-limit headroom from \texttt{os\_proc\_available\_memory}, received
memory warnings, and observed scene interruptions. Failed readings remain
explicit; missing readings are not zero memory.

Physical footprint, RSS, and MLX allocation counters are different metrics
and must not be summed. Our maxima are \textbf{observed sample maxima}, not continuous
kernel high-water marks. Headroom is not system free RAM and does not
guarantee protection from jetsam. A missing warning is not evidence of
no system-wide memory pressure.

The phone runner uses a two-second stable-nominal gate with a five-minute
wait limit. It keeps the display from automatic idle locking during work.
Manual locking, background suspension, auto-brightness, and thermal dimming
remain possible. Four-level thermal labels do not prove identical clock
frequencies or temperatures. The final memory protocol stops on any
non-nominal observed state, warning, interruption, or validation failure;
it does not retry or replace rows automatically.

\hypertarget{ai-assisted-review-and-editing}{%
\subsection{AI-assisted review and editing}\label{ai-assisted-review-and-editing}}

GitHub Copilot and ChatGPT were used in a limited advisory role to review selected system-design details and to assist with manuscript cleanup. Research and experimental decisions remained human-controlled, and suggestions were manually reviewed. AI tools did not autonomously execute experiments, select or discard runs, generate measured data, or determine reported conclusions. All numerical claims are derived from the versioned evidence and deterministic validation scripts described in this paper.

\hypertarget{results}{%
\section{Results}\label{results}}

\hypertarget{the-lru-capacity-cliff-is-not-a-universal-ram-threshold}{%
\subsection{The LRU capacity cliff is not a universal RAM threshold}\label{the-lru-capacity-cliff-is-not-a-universal-ram-threshold}}

The initial short-prompt calibration found substantially fewer reads and shorter
elapsed time when moving from 512 to 576 MiB LRU. The cold-cache medians,
each over five repetitions, are retained here to show the narrow basis
for the current default:

\begin{longtable}[]{@{}
  >{\raggedright\arraybackslash}p{(\columnwidth - 6\tabcolsep) * \real{0.2500}}
  >{\raggedright\arraybackslash}p{(\columnwidth - 6\tabcolsep) * \real{0.2500}}
  >{\raggedright\arraybackslash}p{(\columnwidth - 6\tabcolsep) * \real{0.2500}}
  >{\raggedright\arraybackslash}p{(\columnwidth - 6\tabcolsep) * \real{0.2500}}@{}}
\toprule\noalign{}
\begin{minipage}[b]{\linewidth}\raggedright
Budget (MiB)
\end{minipage} & \begin{minipage}[b]{\linewidth}\raggedright
Cold median elapsed (s)
\end{minipage} & \begin{minipage}[b]{\linewidth}\raggedright
Cold median decode (tokens/s)
\end{minipage} & \begin{minipage}[b]{\linewidth}\raggedright
Cold logical reads (GB)
\end{minipage} \\
\midrule\noalign{}
\endhead
\bottomrule\noalign{}
\endlastfoot
512 & 13.585 & 2.246 & 16.987 \\
576 & 10.810 & 3.052 & 9.709 \\
640 & 11.185 & 2.957 & 9.709 \\
768 & 11.289 & 2.949 & 9.564 \\
1024 & 11.137 & 3.173 & 8.025 \\
\end{longtable}

This result is not a broad optimal-budget finding. On the five later
128-token route captures, the fixed-policy replay gives:

\begin{longtable}[]{@{}llll@{}}
\toprule\noalign{}
Policy & 512 MiB hits (\%) & 576 MiB hits (\%) & 1,024 MiB hits (\%) \\
\midrule\noalign{}
\endhead
\bottomrule\noalign{}
\endlastfoot
LRU & 0.00 & 38.58 & 47.59 \\
FIFO & 0.00 & 25.66 & 40.38 \\
LFU (per residency) & 0.00 & 1.82 & 7.84 \\
Recency-frequency hybrid & 0.00 & 14.65 & 49.83 \\
Random (five-seed mean) & 18.80 & 21.80 & 37.47 \\
Pinned farthest-next-use & 54.49 & 56.93 & 68.23 \\
Relaxed unpinned oracle & 54.73 & 57.13 & 68.31 \\
\end{longtable}

\begin{figure}
\centering
\includegraphics{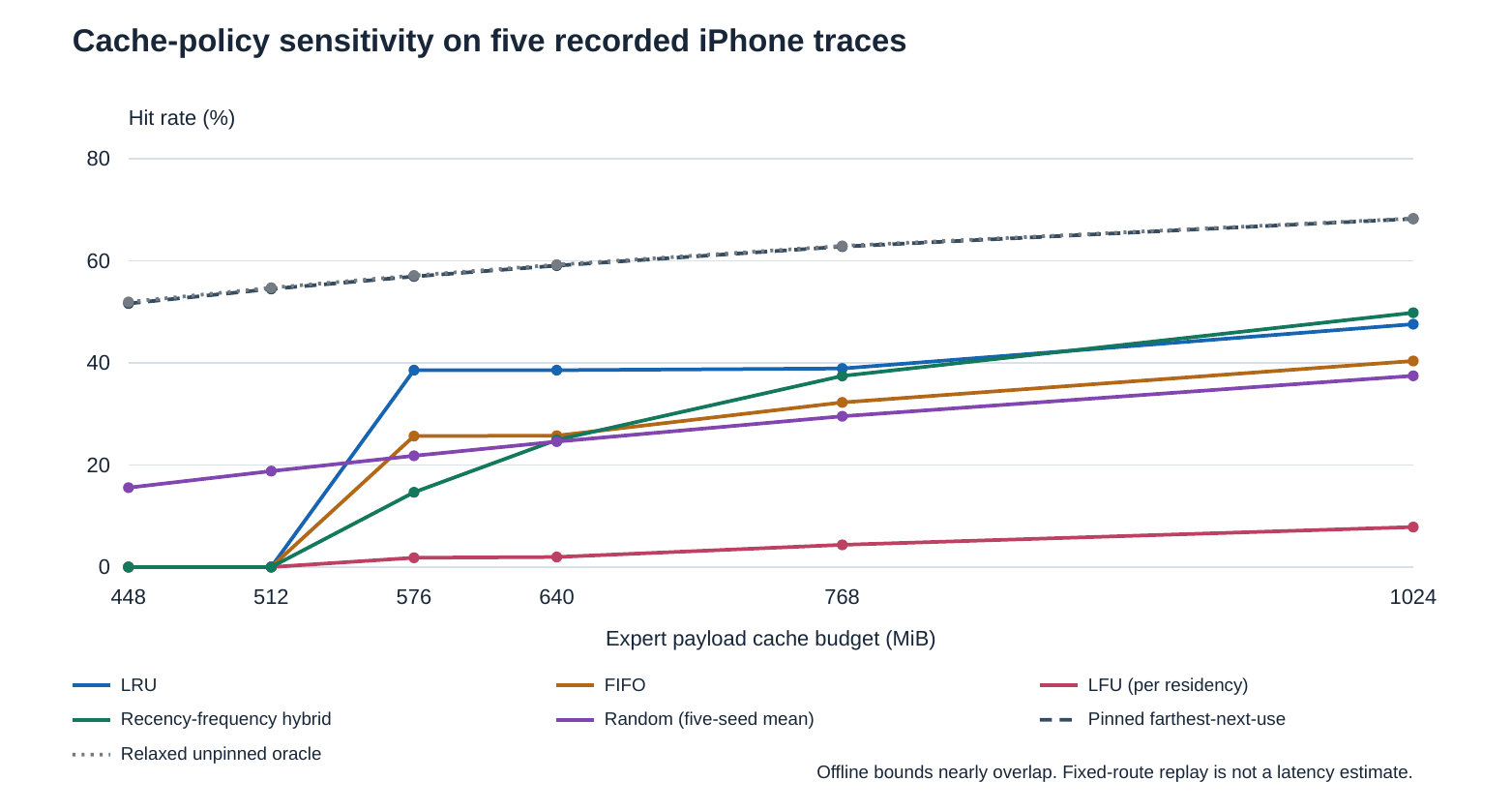}
\caption{Hit rate by payload cache budget; online policies and offline references.}
\end{figure}

\textbf{Figure 1.} Every declared budget is shown. The random band is the observed
seed range, not a confidence interval. The pinned farthest-next-use curve
is a feasible future-aware schedule, not a proven optimum under route
pins. The relaxed unpinned oracle is a lower miss bound under equal-size,
mandatory-admission paging, not an online speed prediction.

The minimum observed distinct-expert reuse distance is 312 on every trace.
At 512 MiB, only 303 whole expert payloads fit; at 576 MiB, 341 fit.
An expert revisited in a later forward has been separated from its last
use by experts from the other 39 layers. The independently computed
reuse histogram predicts every recorded LRU hit count. At least 313 entries
would be needed for any observed LRU reuse on these traces; this is a
necessary access-pattern condition, not a new measured budget or a
whole-process memory requirement.

Seeded random nevertheless attains 18.80\% mean hits at 512 MiB,
with an observed 18.65-18.93\% seed range. Zero reuse is therefore
not unavoidable at that capacity. Policy rankings also change with capacity:
the tested hybrid exceeds LRU at 1,024 MiB. LFU here resets frequency on
eviction; poor performance is not a conclusion about every frequency-based
policy. None of these offline alternatives has a corresponding live phone
speed or energy advantage established by this sweep.

\hypertarget{prediction-accuracy-does-not-imply-useful-prefetch}{%
\subsection{Prediction accuracy does not imply useful prefetch}\label{prediction-accuracy-does-not-imply-useful-prefetch}}

The frozen prefill confidence-0.20 rule predicts an expert present in the
next step's selected set on
5,359/8,741 eligible held-out predictions
(61.31\%). At 512 MiB, the held-out fixed-route replay
adds approximately 1.14\% logical expert bytes. At 576 MiB, the resident-refresh
variant adds 1,120 demand misses and 57 speculative loads, whereas the
preserve-resident variant issues no new loads and exactly reproduces the
no-prefetch demand counters on all five prompts.

The corresponding 576 MiB live counting diagnostic made 1,491 prediction
checks per candidate run, all already resident, yet lost 74 demand hits
through recency changes. Offline replay reproduced the additional
130,940,928 bytes. This is cache pollution caused by a lookup side effect,
not a failed speculative transfer. Zero speculative I/O means an observed
timing difference in that condition cannot demonstrate latency hidden by
overlapping new expert reads.

The additional \texttt{-003} same-runtime Mac matrix exercised actual asynchronous
loads while checking outputs against a demand-filled 20 GiB zero-eviction
reference:

\begin{longtable}[]{@{}
  >{\raggedright\arraybackslash}p{(\columnwidth - 10\tabcolsep) * \real{0.1667}}
  >{\raggedright\arraybackslash}p{(\columnwidth - 10\tabcolsep) * \real{0.1667}}
  >{\raggedright\arraybackslash}p{(\columnwidth - 10\tabcolsep) * \real{0.1667}}
  >{\raggedright\arraybackslash}p{(\columnwidth - 10\tabcolsep) * \real{0.1667}}
  >{\raggedright\arraybackslash}p{(\columnwidth - 10\tabcolsep) * \real{0.1667}}
  >{\raggedright\arraybackslash}p{(\columnwidth - 10\tabcolsep) * \real{0.1667}}@{}}
\toprule\noalign{}
\begin{minipage}[b]{\linewidth}\raggedright
Expert budget (MiB)
\end{minipage} & \begin{minipage}[b]{\linewidth}\raggedright
Prefetch
\end{minipage} & \begin{minipage}[b]{\linewidth}\raggedright
Exact sequences
\end{minipage} & \begin{minipage}[b]{\linewidth}\raggedright
Token comparisons
\end{minipage} & \begin{minipage}[b]{\linewidth}\raggedright
New speculative loads
\end{minipage} & \begin{minipage}[b]{\linewidth}\raggedright
Logical reads (GB)
\end{minipage} \\
\midrule\noalign{}
\endhead
\bottomrule\noalign{}
\endlastfoot
512 & none & 5/5 & 640 & 0 & 556.605 \\
576 & none & 5/5 & 640 & 0 & 348.807 \\
512 & preserve-resident, prefill 0.20 & 5/5 & 640 & 10,334 & 563.879 \\
576 & preserve-resident, prefill 0.20 & 5/5 & 640 & 0 & 348.807 \\
\end{longtable}

The 512 MiB prefetch condition produced 10,334 loads:
6,223 useful and 4,111 wasted blocks, with
1.307\% more logical reads than demand-only execution.
At 576 MiB all those predictions were resident checks and demand counters
matched the control. This is useful coverage of output transparency, not
a measured phone speedup. No demand-to-in-flight-prefetch joins were
observed in this matrix, so it does not cover every asynchronous schedule.

\hypertarget{cache-transparency-is-not-cross-runtime-equivalence}{%
\subsection{Cache transparency is not cross-runtime equivalence}\label{cache-transparency-is-not-cross-runtime-equivalence}}

A prior same-executable Swift control compared 576 MiB against a 20 GiB
demand-filled cache on five \texttt{-001} prompts: all 640 output tokens matched,
despite 183,699 evictions in the smaller cache and none in the larger one.
The \texttt{-003} matrix adds 20 exact 128-token comparisons, or 2,560 token
comparisons, covering bounded demand caches and active prefetch. Across
the two designs this is ten distinct fixed prompts, not an exhaustive
correctness proof. The large-cache reference is not a preloaded upstream
resident model.

The resident-Python comparison is a different experiment. It reuses exact
prepared phone inputs, serial prefill, and greedy selection under pinned
MLX/MLX-Metal 0.31.2. No complete 128-token sequence matches the archived
phone output:

\begin{longtable}[]{@{}
  >{\raggedright\arraybackslash}p{(\columnwidth - 4\tabcolsep) * \real{0.3333}}
  >{\raggedright\arraybackslash}p{(\columnwidth - 4\tabcolsep) * \real{0.3333}}
  >{\raggedright\arraybackslash}p{(\columnwidth - 4\tabcolsep) * \real{0.3333}}@{}}
\toprule\noalign{}
\begin{minipage}[b]{\linewidth}\raggedright
Prompt
\end{minipage} & \begin{minipage}[b]{\linewidth}\raggedright
First different output position (1-based)
\end{minipage} & \begin{minipage}[b]{\linewidth}\raggedright
Fixed-history next-token matches
\end{minipage} \\
\midrule\noalign{}
\endhead
\bottomrule\noalign{}
\endlastfoot
conversation-001 & 3 & 123/128 \\
code-001 & 83 & 127/128 \\
mathematics-001 & 73 & 126/128 \\
reasoning-001 & 2 & 127/128 \\
expository-001 & 1 & 121/128 \\
\end{longtable}

Feeding back the recorded phone history yields
624/640 matching next-token predictions
(97.5\%). This is \textbf{conditional next-token agreement}, not
free-generation accuracy, task quality, or evidence that only a comparable
fraction of a free-running response is wrong. The official upstream
generator reproduces the custom resident loop's outputs, ruling out that
loop as the source of the observed sequence discrepancy.

A raw-token-0 diagnostic further shows that changing only Python
\texttt{mlx}/\texttt{mlx-metal} from 0.32.2 to the supported minimum 0.31.2 changes the
greedy token from 951 to 198 and recovers the saved phone fingerprints;
reversion restores the original result. The unsupported 0.31.1 Python
control was blocked before inference, not forced into execution. Fingerprint
agreement is at the recorded field precision, not full-vector equivalence.
A subsequent two-token state probe localized an arithmetic difference
and recovered several complete tensors with an opt-in native convolution,
but whole-model native modes still did not establish sequence equivalence.
The phone arithmetic defaults were not promoted on the basis of a
component-level match.

\hypertarget{sustained-execution-and-timing-variation}{%
\subsection{Sustained execution and timing variation}\label{sustained-execution-and-timing-variation}}

The four manually sequenced 512-token counting runs all reached their cap
with nominal app-sampled peak and final thermals:

\begin{longtable}[]{@{}
  >{\raggedright\arraybackslash}p{(\columnwidth - 10\tabcolsep) * \real{0.1667}}
  >{\raggedright\arraybackslash}p{(\columnwidth - 10\tabcolsep) * \real{0.1667}}
  >{\raggedright\arraybackslash}p{(\columnwidth - 10\tabcolsep) * \real{0.1667}}
  >{\raggedright\arraybackslash}p{(\columnwidth - 10\tabcolsep) * \real{0.1667}}
  >{\raggedright\arraybackslash}p{(\columnwidth - 10\tabcolsep) * \real{0.1667}}
  >{\raggedright\arraybackslash}p{(\columnwidth - 10\tabcolsep) * \real{0.1667}}@{}}
\toprule\noalign{}
\begin{minipage}[b]{\linewidth}\raggedright
Order
\end{minipage} & \begin{minipage}[b]{\linewidth}\raggedright
Policy
\end{minipage} & \begin{minipage}[b]{\linewidth}\raggedright
Elapsed (s)
\end{minipage} & \begin{minipage}[b]{\linewidth}\raggedright
Decode (tokens/s)
\end{minipage} & \begin{minipage}[b]{\linewidth}\raggedright
Logical reads (GB)
\end{minipage} & \begin{minipage}[b]{\linewidth}\raggedright
Peak thermal
\end{minipage} \\
\midrule\noalign{}
\endhead
\bottomrule\noalign{}
\endlastfoot
A1 & None & 314.105 & 1.795 & 319.354 & nominal \\
B1 & Prefill 0.20 & 272.711 & 2.085 & 320.297 & nominal \\
B2 & Prefill 0.20 & 301.553 & 1.864 & 320.297 & nominal \\
A2 & None & 304.148 & 1.852 & 319.354 & nominal \\
\end{longtable}

The two identically configured candidate runs differ by
10.58\% elapsed time. Nominal thermal labels therefore
do not remove temporal performance variation. The larger apparent win
in the first pair does not establish a stable policy effect. These records
show sustained-generation feasibility on one workload; their older
MLX peaks are not process-footprint measurements, and lifecycle counts
were not present in that single-run schema.

The later iOS 27 cache-budget campaign preserves the following paired
elapsed comparisons, including its stopping row:

\begin{longtable}[]{@{}
  >{\raggedright\arraybackslash}p{(\columnwidth - 10\tabcolsep) * \real{0.1667}}
  >{\raggedright\arraybackslash}p{(\columnwidth - 10\tabcolsep) * \real{0.1667}}
  >{\raggedright\arraybackslash}p{(\columnwidth - 10\tabcolsep) * \real{0.1667}}
  >{\raggedright\arraybackslash}p{(\columnwidth - 10\tabcolsep) * \real{0.1667}}
  >{\raggedright\arraybackslash}p{(\columnwidth - 10\tabcolsep) * \real{0.1667}}
  >{\raggedright\arraybackslash}p{(\columnwidth - 10\tabcolsep) * \real{0.1667}}@{}}
\toprule\noalign{}
\begin{minipage}[b]{\linewidth}\raggedright
Prompt / pair
\end{minipage} & \begin{minipage}[b]{\linewidth}\raggedright
512 MiB elapsed (s)
\end{minipage} & \begin{minipage}[b]{\linewidth}\raggedright
576 MiB elapsed (s)
\end{minipage} & \begin{minipage}[b]{\linewidth}\raggedright
576 vs 512 change (\%)
\end{minipage} & \begin{minipage}[b]{\linewidth}\raggedright
Read reduction (\%)
\end{minipage} & \begin{minipage}[b]{\linewidth}\raggedright
Both nominal?
\end{minipage} \\
\midrule\noalign{}
\endhead
\bottomrule\noalign{}
\endlastfoot
conversation-001 / 1 & 83.782 & 74.811 & -10.71 & 40.10 & yes \\
conversation-001 / 2 & 75.045 & 62.730 & -16.41 & 40.10 & yes \\
code-001 / 1 & 84.173 & 70.838 & -15.84 & 38.16 & yes \\
code-001 / 2 & 76.458 & 81.946 & +7.18 & 38.16 & yes \\
mathematics-001 / 1 & 89.128 & 82.120 & -7.86 & 39.01 & yes \\
mathematics-001 / 2 & 86.120 & 76.707 & -10.93 & 39.01 & NO; retained stop row \\
\end{longtable}

Elapsed here follows the saved generation scope and excludes separately
recorded drain. The second fully nominal code pair is slower at 576 MiB,
despite lower logical reads. The second mathematics pair includes a
fair-thermal run and is \textbf{not a valid nominal comparison}. We keep it
visible rather than discard it or quietly replace it.

\hypertarget{sampled-process-memory-and-a-retained-thermal-stop}{%
\subsection{Sampled process memory and a retained thermal stop}\label{sampled-process-memory-and-a-retained-thermal-stop}}

The predeclared parent protocol called for 14 load-inclusive requests.
Twelve completed their 128-token generations; the first eleven passed
all conditions. Request 12 finished with fair final/peak thermal state,
triggering the required stop. All twelve outputs and memory reports
remain saved. This was neither a crash nor a memory-warning event, and
the thermal label alone does not prove throttling.

After inspecting that stop, we separately declared only the two previously
unexecuted longer-context requests. They retain the same text, eight
context repetitions, 256-512 prepared-token bounds, 32-token cap, and
576-then-512 MiB order. Both actually used
340 prepared tokens and produced identical 32-token
outputs, including the sampled end token. Both passed every original
condition. They do not amend the failed parent into a completed nominal
14-run study.

\begin{longtable}[]{@{}
  >{\raggedright\arraybackslash}p{(\columnwidth - 8\tabcolsep) * \real{0.2000}}
  >{\raggedright\arraybackslash}p{(\columnwidth - 8\tabcolsep) * \real{0.2000}}
  >{\raggedright\arraybackslash}p{(\columnwidth - 8\tabcolsep) * \real{0.2000}}
  >{\raggedright\arraybackslash}p{(\columnwidth - 8\tabcolsep) * \real{0.2000}}
  >{\raggedright\arraybackslash}p{(\columnwidth - 8\tabcolsep) * \real{0.2000}}@{}}
\toprule\noalign{}
\begin{minipage}[b]{\linewidth}\raggedright
Cohort / cache budget
\end{minipage} & \begin{minipage}[b]{\linewidth}\raggedright
Prompt/output tokens
\end{minipage} & \begin{minipage}[b]{\linewidth}\raggedright
Requests
\end{minipage} & \begin{minipage}[b]{\linewidth}\raggedright
Sampled footprint peaks (GiB)
\end{minipage} & \begin{minipage}[b]{\linewidth}\raggedright
Peak thermals
\end{minipage} \\
\midrule\noalign{}
\endhead
\bottomrule\noalign{}
\endlastfoot
Stopped parent / 512 and 576 MiB & 49-69/128 & 12 & 1.87-2.32 & 11 nominal; 1 fair (stop) \\
Separate follow-up / 576 MiB & 340/32 & 1 & 2.39 & nominal \\
Separate follow-up / 512 MiB & 340/32 & 1 & 2.73 & nominal \\
\end{longtable}

Appendix B retains every request's baseline, footprint peak, RSS peak,
budget, and thermal status; no stopping row is removed from that record.

\begin{figure}
\centering
\includegraphics{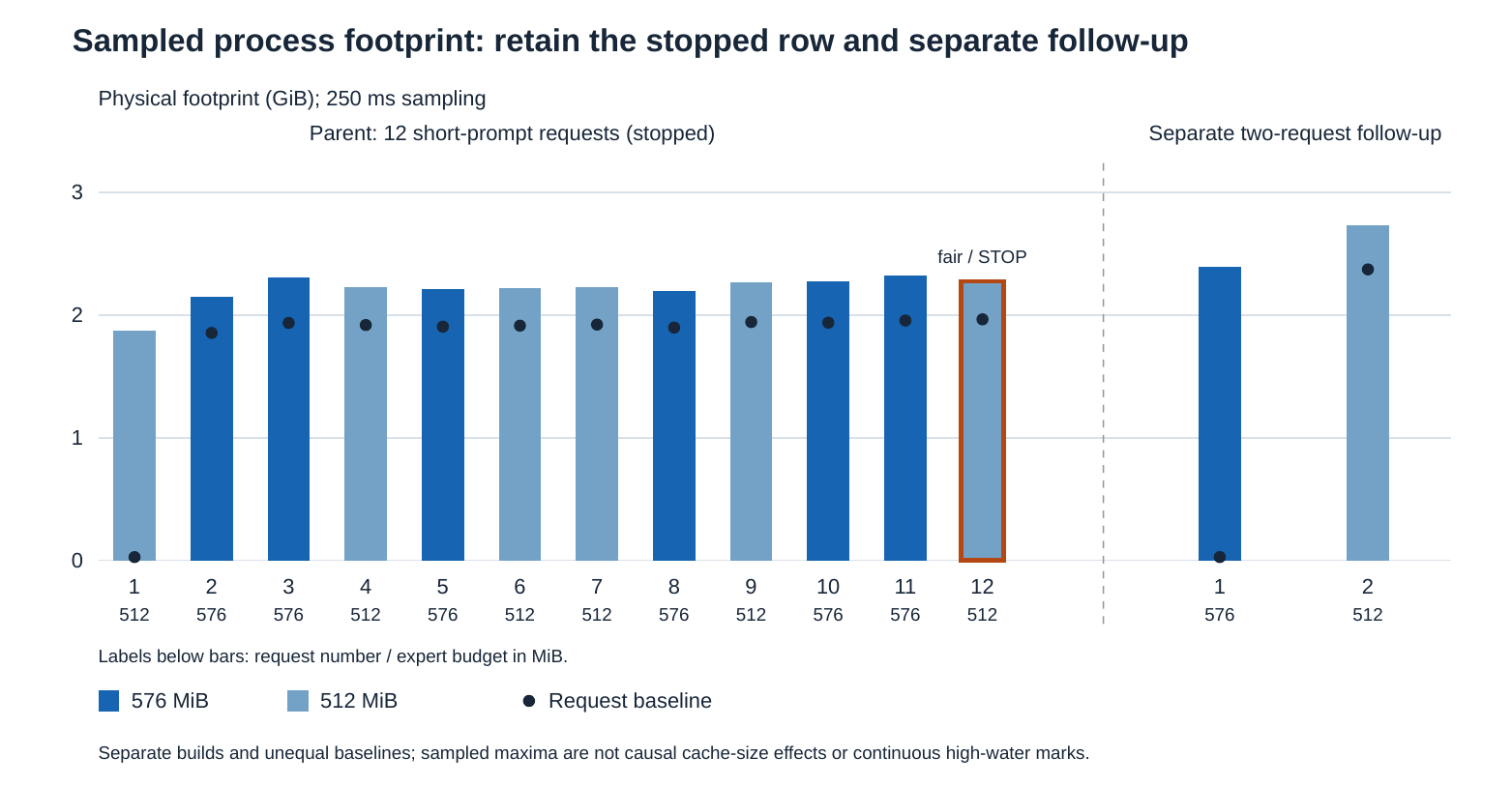}
\caption{All sampled process-footprint peaks and start readings, with the parent stopping row retained and the follow-up separated.}
\end{figure}

\textbf{Figure 2.} Bars are sampled per-request physical-footprint peaks; dots
are start readings. The parent and follow-up have different executable
digests. Within each process, model recreation does not empty allocator
or filesystem caches. In particular, the second longer-context request's
much larger starting footprint makes a causal cache-size memory comparison
invalid.

The parent contains 3,793 successful memory samples and
the follow-up 1,647, with no sampling failures, received
memory warnings, or observed scene interruptions. The largest
successful-sample gap across both is 0.255136 seconds. Short-prompt
sampled footprint peaks span 1.87-2.32 GiB.
The fixed longer prompt reaches 2.39 GiB in the first
request and 2.73 GiB in the second. These observations
address the prior absence of whole-process measurements for the tested
workloads. They do not establish a context-length scaling curve or a
device-wide memory bound.

\hypertarget{a-qualified-whole-device-power-case-study}{%
\subsection{A qualified whole-device power case study}\label{a-qualified-whole-device-power-case-study}}

One standalone iOS Performance Trace recording contains a complete
512-token counting request under the confidence-gated prefetch policy.
The request-plus-drain wall-clock window is 263.382 seconds.
Reconstructing original SystemMetrics Begin-End intervals gives nearly
complete coverage, no observed charging overlap, a duration-weighted
26.505 battery-percent/hour rate, and a
1.939-percentage-point \textbf{source-window rate-time estimate}.

This is the integral of a recorded whole-device estimator over the aligned
window. It is not app-only watts or joules, a direct battery-gauge delta,
or calibrated energy per token. Display brightness varied from 22\% to
44\%, the source estimator is not independently calibrated here, and there
is no matched no-prefetch power control. The recording supports a narrowly
described power case study, not an energy-saving claim. The underlying
profiling archive is withheld from the source tree because it may contain
unrelated device metadata; the scoped derived record is retained.

\hypertarget{discussion-and-threats-to-validity}{%
\section{Discussion and threats to validity}\label{discussion-and-threats-to-validity}}

\textbf{Feasibility is not optimality.} The implementation executes a large
quantized checkpoint's text path on a physical phone, but that does not
make it the best use of the device's memory, storage, latency budget, or
battery. There is no completed matched study establishing superiority
over smaller resident 4B/8B models.

\textbf{Serial prefill is a major implementation boundary.} Prompt tokens are
processed one at a time with full output-head evaluation. This shapes
TTFT, reuse, and expert traffic. The measured 340-token prompt is longer
than the short prompts in these protocols, not ``long context'' in the
sense of thousands of tokens. Batched prefill or a different prompt mix
could change both cache ranking and latency.

\textbf{Memory and I/O scopes are limited.} Expert payload budgets exclude
other allocations. Process footprint is sampled, not a complete record
of all instantaneous allocations or system filesystem caches. Logical
expert reads omit other runtime reads and do not expose physical flash
behavior. We cannot infer NAND bandwidth or storage endurance from the
large logical totals.

\textbf{Correctness depends on the comparison boundary.} The most defensible
transparency claim holds the Swift runtime, arithmetic, and hardware
fixed while varying cache behavior. Cross-framework and cross-device
sequence equivalence remains unestablished. Prediction precision and
conditional token agreement are not task-quality metrics. No lossy
routing modification or quality benefit is evaluated.

\textbf{Small, correlated workloads.} Five replay traces, five additional
Mac prompts, and one phone provide mechanisms and counterexamples, not
population estimates. Multiple seeds and budgets reuse the same traces.
Greedy generation improves repeatability but does not cover sampling-based
decoding, concurrent users, batching, all context lengths, or background
application conditions.

\textbf{Failure and scheduling coverage is incomplete.} The cache and sampler
have focused contract tests, but the final live prefetch matrix observed
no demand joins or expert-load failures. It cannot establish behavior
under every cancellation, I/O failure, scheduling race, or memory-pressure
path. A received-warning count of zero does not establish the absence
of every possible termination risk.

\textbf{Evolving instrumentation.} Device OS versions, compilers, dependency
versions, and export schemas changed across this multi-stage study.
We do not pool timings across these changes. Older output exports cannot
acquire missing provenance retrospectively: a balanced-export bug could
attach the warm-up generation object to measured schema-6 rows. We do
not use those legacy output objects as independent token-equivalence
evidence; the later raw same-runtime matrices and schema-12 captures
provide the stronger comparisons.

\textbf{Reproducibility and artifact scope.} The checkpoint, pinned
dependencies, format, protocols, evidence hashes, and source snapshots are
versioned in the research archive. The public runtime, analysis code, and
paper-facing artifact are available in the
\href{https://github.com/MusaShams/Routide}{Routide GitHub repository}. The
public preprint source contains the manuscript and figures needed to reproduce
this document, while licensed model weights and private profiling archives
remain outside the paper package. Full system reproduction additionally
requires the runtime, packing workflow, model weights under their applicable
license, and suitable Apple hardware.

\hypertarget{conclusion}{%
\section{Conclusion}\label{conclusion}}

Routide demonstrates and characterizes storage-backed execution of a
pinned large MoE checkpoint on an iPhone. Its evidence supports more
than a text-generation demo but less than a universal efficiency claim.
The cache cliff is policy-dependent; resident-prefetch checks can damage
reuse without doing useful I/O; exact same-runtime cache transparency
does not imply cross-runtime sequence equivalence; and process-memory
measurements require their own scopes rather than relabeled allocator
counters. Preserving negative pairs, a thermal stop, and a separately
declared successful continuation makes these distinctions testable.
These findings motivate independent reproduction and careful comparison
boundaries rather than repeated benchmarks selected for favorable timing.

\hypertarget{references}{%
\section*{References}}

{[}1{]} MLX Community. \emph{Qwen3.6-35B-A3B-4bit}.
\href{https://huggingface.co/mlx-community/Qwen3.6-35B-A3B-4bit/tree/38740b847e4cb78f352aba30aa41c76e08e6eb46}{Pinned model repository}.
Upstream model: \href{https://huggingface.co/Qwen/Qwen3.6-35B-A3B}{Qwen/Qwen3.6-35B-A3B}.
The retrieved community model card declares Apache-2.0; retain the applicable
upstream license when distributing weights.

{[}2{]} Keivan Alizadeh et al.~\emph{LLM in a flash: Efficient Large Language Model
Inference with Limited Memory}. arXiv:2312.11514, 2023; revised 2024.
\href{https://arxiv.org/abs/2312.11514}{Abstract and version history};
\href{https://arxiv.org/html/2312.11514v3}{version 3}.

{[}3{]} Zhenliang Xue, Yixin Song, Zeyu Mi, Xinrui Zheng, Yubin Xia, and Haibo Chen.
\emph{PowerInfer-2: Fast Large Language Model Inference on a Smartphone}.
arXiv:2406.06282, 2024, version 3.
\href{https://arxiv.org/abs/2406.06282v3}{Paper}.

{[}4{]} Keisuke Kamahori, Yile Gu, Kan Zhu, and Baris Kasikci.
\emph{Fiddler: CPU-GPU Orchestration for Fast Inference of Mixture-of-Experts
Models}. arXiv:2402.07033, 2024; version 3 revised 2025.
\href{https://arxiv.org/abs/2402.07033v3}{Paper}.

{[}5{]} MLX contributors. \href{https://github.com/ml-explore/mlx-swift}{MLX Swift}
and \href{https://github.com/ml-explore/mlx-swift-examples/tree/378f2449c257788c5067b9f8b086731d76b39b33}{MLX Swift examples}.
Routide derives from the latter pinned example revision; MIT license and
upstream acknowledgments are retained.

{[}6{]} L. A. Belady. \emph{A study of replacement algorithms for a virtual-storage
computer}. IBM Systems Journal, 5(2), 78-101, 1966.
\href{https://doi.org/10.1147/sj.52.0078}{DOI}.

\appendix

\hypertarget{claim-boundaries}{%
\section{Claim boundaries}\label{claim-boundaries}}

\begin{longtable}[]{@{}
  >{\raggedright\arraybackslash}p{(\columnwidth - 2\tabcolsep) * \real{0.5000}}
  >{\raggedright\arraybackslash}p{(\columnwidth - 2\tabcolsep) * \real{0.5000}}@{}}
\toprule\noalign{}
\begin{minipage}[b]{\linewidth}\raggedright
Statement supported here
\end{minipage} & \begin{minipage}[b]{\linewidth}\raggedright
Stronger statement not established
\end{minipage} \\
\midrule\noalign{}
\endhead
\bottomrule\noalign{}
\endlastfoot
Text generation from the pinned flash-paged model on one iPhone & Universal device/model feasibility or state-of-the-art speed \\
Policy-dependent hits on five recorded traces & An unavoidable minimum RAM threshold or globally optimal 576 MiB budget \\
Exact tested outputs with the same Swift implementation under cache changes & General cross-framework or cross-device numerical equivalence \\
A fixed predictor's selected-set overlap and read accounting & General latency hiding or energy savings \\
Sampled process footprint on separately identified protocols & Continuous device-wide memory maxima or context-length scaling \\
One whole-device power-estimator integral & Calibrated joules/token or app-only energy benefit \\
A separate successful longer-prompt pair after a retained stop & A completed nominal 14-request parent campaign \\
Structured prompts and captured text & Scored task-quality parity or superiority to smaller resident models \\
\end{longtable}

These boundaries define the claims supported by the present evidence.
Stronger claims require additional experiments or independent reproduction.

\hypertarget{per-request-process-memory-observations}{%
\section{Per-request process-memory observations}\label{per-request-process-memory-observations}}

Every completed observation is retained below. The parent study stopped
after its twelfth request; the two follow-up rows belong to a separately
declared protocol and a different executable. Footprint and RSS columns
are separate sampled maxima, not additive memory components.

\begin{longtable}[]{@{}
  >{\raggedright\arraybackslash}p{(\columnwidth - 12\tabcolsep) * \real{0.1429}}
  >{\raggedright\arraybackslash}p{(\columnwidth - 12\tabcolsep) * \real{0.1429}}
  >{\raggedright\arraybackslash}p{(\columnwidth - 12\tabcolsep) * \real{0.1429}}
  >{\raggedright\arraybackslash}p{(\columnwidth - 12\tabcolsep) * \real{0.1429}}
  >{\raggedright\arraybackslash}p{(\columnwidth - 12\tabcolsep) * \real{0.1429}}
  >{\raggedright\arraybackslash}p{(\columnwidth - 12\tabcolsep) * \real{0.1429}}
  >{\raggedright\arraybackslash}p{(\columnwidth - 12\tabcolsep) * \real{0.1429}}@{}}
\toprule\noalign{}
\begin{minipage}[b]{\linewidth}\raggedright
Protocol/run
\end{minipage} & \begin{minipage}[b]{\linewidth}\raggedright
Cache (MiB)
\end{minipage} & \begin{minipage}[b]{\linewidth}\raggedright
Prompt/output tokens
\end{minipage} & \begin{minipage}[b]{\linewidth}\raggedright
Start footprint (GiB)
\end{minipage} & \begin{minipage}[b]{\linewidth}\raggedright
Peak footprint (GiB)
\end{minipage} & \begin{minipage}[b]{\linewidth}\raggedright
Peak RSS (GiB)
\end{minipage} & \begin{minipage}[b]{\linewidth}\raggedright
Peak thermal
\end{minipage} \\
\midrule\noalign{}
\endhead
\bottomrule\noalign{}
\endlastfoot
Parent/1 & 512 & 49/128 & 0.027 & 1.874 & 1.770 & nominal \\
Parent/2 & 576 & 49/128 & 1.854 & 2.152 & 2.070 & nominal \\
Parent/3 & 576 & 49/128 & 1.935 & 2.305 & 2.104 & nominal \\
Parent/4 & 512 & 49/128 & 1.918 & 2.226 & 2.102 & nominal \\
Parent/5 & 576 & 52/128 & 1.905 & 2.212 & 2.100 & nominal \\
Parent/6 & 512 & 52/128 & 1.913 & 2.221 & 2.102 & nominal \\
Parent/7 & 512 & 52/128 & 1.922 & 2.226 & 2.108 & nominal \\
Parent/8 & 576 & 52/128 & 1.897 & 2.193 & 2.094 & nominal \\
Parent/9 & 512 & 69/128 & 1.942 & 2.263 & 2.089 & nominal \\
Parent/10 & 576 & 69/128 & 1.937 & 2.271 & 2.115 & nominal \\
Parent/11 & 576 & 69/128 & 1.954 & 2.318 & 2.117 & nominal \\
Parent/12 & 512 & 69/128 & 1.964 & 2.274 & 2.118 & fair; STOP \\
Follow-up/1 & 576 & 340/32 & 0.028 & 2.390 & 1.815 & nominal \\
Follow-up/2 & 512 & 340/32 & 2.371 & 2.729 & 2.104 & nominal \\
\end{longtable}

\end{document}